\documentclass[journal=ancham,manuscript=article,keywords=true]{achemso}

\usepackage{graphicx}
\usepackage{url}
\usepackage{booktabs}
\usepackage{xcolor}
\definecolor{revisionpurple}{RGB}{128,0,128} 
\definecolor{commentorange}{RGB}{230,120,0} 
\definecolor{fillgreen}{RGB}{0,140,0}      
\usepackage{amsmath}
\usepackage{array}       
\newcolumntype{L}[1]{>{\raggedright\arraybackslash}p{#1}}
\newcolumntype{C}[1]{>{\centering\arraybackslash}p{#1}}

\newcommand{\fittable}[1]{\resizebox{\textwidth}{!}{#1}}

\graphicspath{{figures/}}

\title{Simulation-Supervised Foundation Models for Retention Time Prediction in High-Performance Liquid Chromatography beyond Experimental Data Coverage}

\author{Stephen Wu}
\affiliation{The Institute of Statistical Mathematics, Research Organization of Information and Systems, 10-3 Midori-cho, Tachikawa, Tokyo 190-8562, Japan}
\alsoaffiliation{Graduate Institute for Advanced Studies, SOKENDAI, 10-3 Midori-cho, Tachikawa, Tokyo 190-8562, Japan}

\email{stewu@ism.ac.jp}

\author{Yufeng Han}
\affiliation{City University of Hong Kong, 83 Tat Chee Avenue, Kowloon, Hong Kong SAR, China}

\author{Yasuhiro Mito}
\affiliation{Research Center, Shimadzu General Services, Inc., 1 Nishinokyo-Kuwabara-cho, Nakagyo-ku, Kyoto 604-8511, Japan}

\author{Yoshiyuki Watabe}
\affiliation{Research Center, Shimadzu General Services, Inc., 1 Nishinokyo-Kuwabara-cho, Nakagyo-ku, Kyoto 604-8511, Japan}

\author{Yoshihiro Hayashi}
\affiliation{The Institute of Statistical Mathematics, Research Organization of Information and Systems, 10-3 Midori-cho, Tachikawa, Tokyo 190-8562, Japan}
\alsoaffiliation{Advanced General Intelligence for Science Program (AGIS), TRIP Headquarters, RIKEN, 2-1 Hirosawa, Wako, Saitama 351-0198, Japan}
\alsoaffiliation{Graduate Institute for Advanced Studies, SOKENDAI, 10-3 Midori-cho, Tachikawa, Tokyo 190-8562, Japan}

\author{Hikaru Takaya}
\affiliation{Department of Life and Health Sciences, Faculty of Life and Environmental Sciences, Teikyo University of Science, 2-2-1 Senju-Sakuragi, Adachi-ku, Tokyo 120-0045, Japan}

\author{Takuya Kubo}
\affiliation{Division of Applied Life Sciences, Graduate School of Life and Environmental Sciences, Kyoto Prefectural University, 1-5 Shimogamo Hangi-cho, Sakyo-ku, Kyoto 606-8522, Japan}

\author{Ryo Yoshida}
\affiliation{The Institute of Statistical Mathematics, Research Organization of Information and Systems, 10-3 Midori-cho, Tachikawa, Tokyo 190-8562, Japan}
\alsoaffiliation{Advanced General Intelligence for Science Program (AGIS), TRIP Headquarters, RIKEN, 2-1 Hirosawa, Wako, Saitama 351-0198, Japan}
\alsoaffiliation{Graduate Institute for Advanced Studies, SOKENDAI, 10-3 Midori-cho, Tachikawa, Tokyo 190-8562, Japan}

\email{yoshidar@ism.ac.jp}


\begin{document}

\begin{tocentry}
	\centering
	\includegraphics[width=\linewidth]{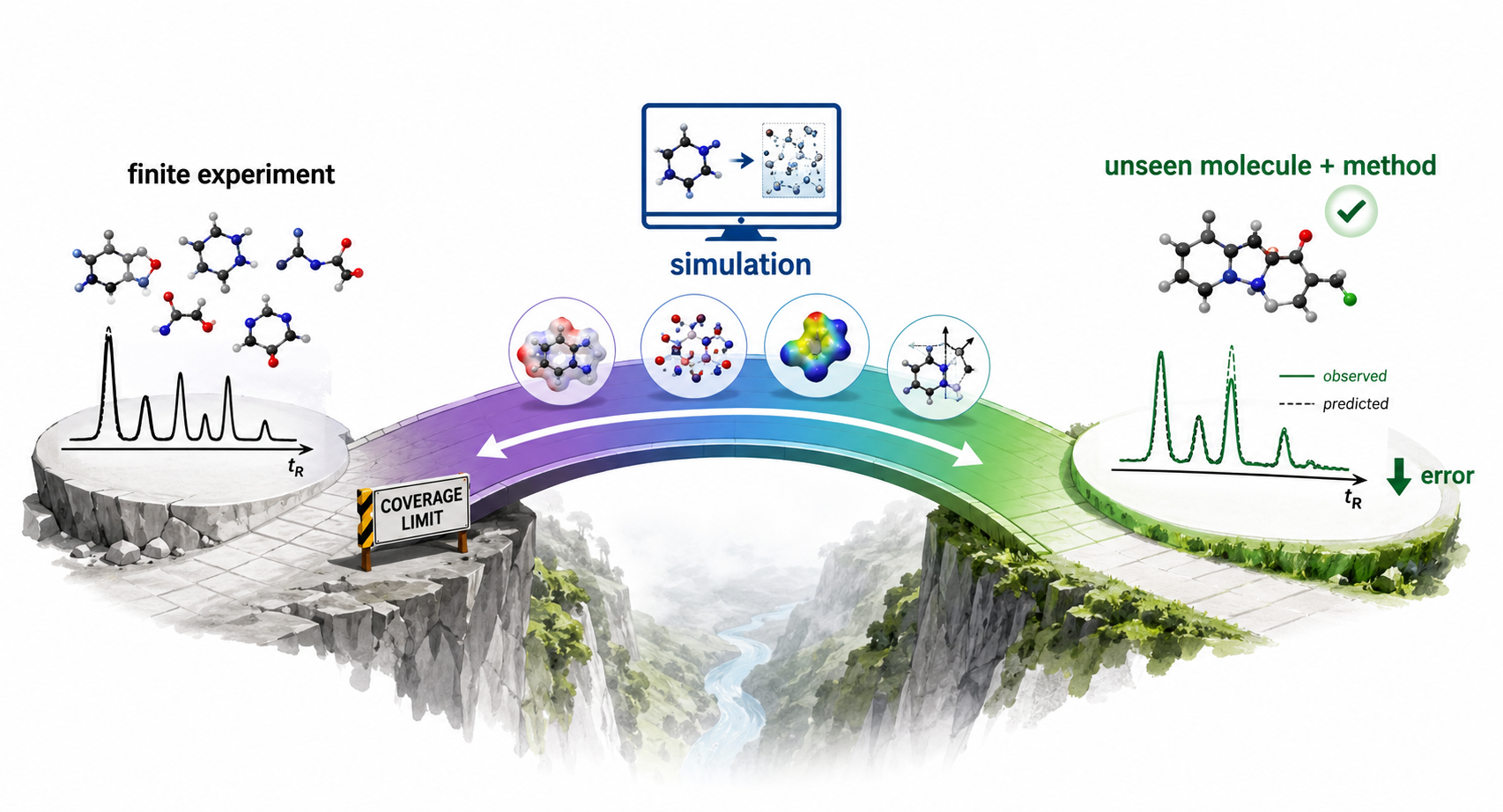}
\end{tocentry}

\begin{abstract}

Accurate prediction of high-performance liquid chromatography (HPLC) retention times (RTs) across diverse molecules and chromatographic methods remains challenging because experimental training data cover only a limited region of chemical and method spaces. Here, we develop \mbox{FUSE-RT} (\emph{F}oundation model \emph{U}nifying \emph{S}imulation and \emph{E}xperimental supervision for \emph{R}etention \emph{T}ime), a multitask foundation model that integrates RT data from 179 chromatographic methods and adapts to unseen molecules and methods using limited target-domain data. To extend transferability beyond experimental coverage, we introduce simulation-to-real (Sim2Real) transfer learning, in which molecular representations learned from large-scale computational data are transferred to experimental RT prediction. Specifically, we use PolyOmics, comprising 39 properties for approximately 21,400 molecules generated by molecular dynamics and density-functional theory calculations, as auxiliary supervision. We evaluate generalization under molecular, method, and joint molecular--method distribution shifts. Simulation-derived supervision substantially improves transfer beyond the experimental molecular domain, particularly under pronounced coverage gaps and few-shot adaptation. Moreover, RT-prediction error decreases systematically with increasing simulation-data size, following a significant power-law relationship. These results establish Sim2Real transfer as a scalable strategy for extending RT prediction beyond the finite coverage of experimental chromatographic data.

\end{abstract}

\vspace{1em}

\section{Introduction}

Accurate prediction of high-performance liquid chromatography (HPLC) retention times (RTs) is a long-standing challenge in analytical chemistry, with direct implications for method development, compound identification, and high-throughput screening workflows~\cite{heberger2007qsrr,haddad2021retentionreview}. Retention behavior depends strongly on chromatographic conditions, including stationary-phase chemistry, mobile-phase composition, gradient profile, temperature, and instrument-specific factors. Consequently, experimental RT data are intrinsically heterogeneous, and RT-prediction models trained under one set of conditions often generalize poorly to new experimental setups. Because generating RT datasets for every chromatographic configuration is costly, a central challenge is to develop predictive models that generalize across both chemical space and chromatographic conditions.

Machine-learning (ML) methods have increasingly addressed this challenge by leveraging growing collections of chromatographic data such as RepoRT~\cite{report2024}. Early quantitative structure--retention relationship models were typically trained separately for individual chromatographic setups~\cite{heberger2007qsrr}. More recent studies have shifted toward global or multi-condition modeling, including cross-system retention mapping and models that incorporate chromatographic parameters as inputs to predict RTs across diverse chromatographic conditions~\cite{stanstrup2015predret,pasin2021single_rt_model,souihi2022multiconditionrt,li2025multidataset,peirovila2024globalretention2,george2024grip}. In parallel, deep neural networks have enabled learned molecular representations from large RT datasets~\cite{domingoalmenara2019smrt,kensert2021gcn,yang2021gnn_rt,xu2023qgegnn,bouwmeester2021deeplc}, while transfer learning has been used to adapt pretrained representations to new tasks using limited target datasets~\cite{ju2021autoencoder_rt,kwon2023pretrainedgnn,stienstra2025graphormerrt,hong2025tstl,bouwmeester2026deeplc_transfer}. Related work has focused on predicting retention order rather than absolute RT~\cite{bach2018retentionorder}. Recent graph-transformer and multitask architectures have further expanded the range of chromatographic conditions that can be modeled within a shared network~\cite{ying2021graphormer,stienstra2025graphormerrt,xiong2026unirt}. Collectively, these developments reflect a broader trend toward large pretrained and multitask models that learn shared representations transferable across downstream tasks~\cite{bommasani2021foundation,raffel2020t5,ramsundar2015multitask,tripuraneni2020taskdiversity,wu2018moleculenet}.

Here, we develop a multitask foundation model for RT prediction by jointly learning from the RT data of 179 diverse chromatographic methods. The pretrained model can be adapted to previously unseen molecules and chromatographic methods using a limited set of experimentally measured RTs from the target domain, providing a practical route to constructing customized prediction models without collecting laboratory datasets covering a large set of target molecules and chromatographic conditions. The effectiveness of such transfer learning, however, depends strongly on the coverage of the pretraining data. Public resources such as RepoRT provide valuable experimental RT measurements but necessarily cover only a finite range of molecular structures and chromatographic conditions. When target molecules and chromatographic methods differ substantially from those represented during pretraining, the benefits of fine-tuning can diminish, with performance approaching that of training from scratch. Systematically expanding experimental coverage is challenging because acquiring RT measurements across the enormous joint space of molecular structures and chromatographic conditions requires substantial laboratory effort.

To extend model transferability beyond the coverage of experimental data, we investigate large-scale molecular simulation as a scalable source of auxiliary supervision. Using RadonPy~\cite{hayashi2022radonpy}, an automated molecular dynamics (MD) simulation framework, we construct a computational dataset comprising 39 properties for approximately 21,400 structurally diverse molecules and jointly train the model to predict these properties together with RTs across multiple chromatographic methods. The simulated targets---including thermal, optical, electronic, and solubility-related properties---are general molecular and material properties rather than direct descriptors of chromatographic retention. Instead, they provide diverse, physically grounded supervision for learning molecular representations that extend beyond the chemical space covered by experimental RT data.

\begin{figure}[htbp]
\centering
\includegraphics[width=\linewidth]{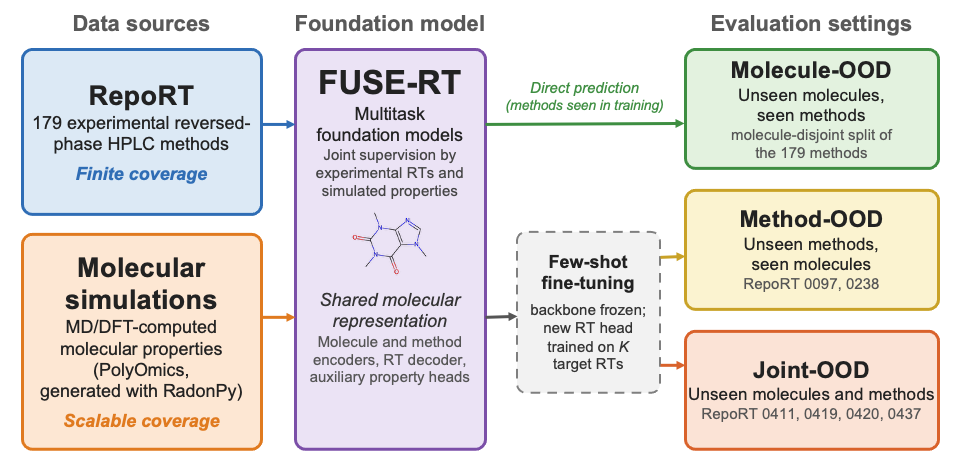}
\caption{\textbf{Overview of the development and evaluation of the RT-prediction foundation models FUSE-RT.} Two complementary data sources---a finite collection of 179 experimental datasets from RepoRT corresponding to different chromatographic methods and the scalable molecular simulation database PolyOmics, generated using RadonPy---jointly supervise FUSE-RT.
For chromatographic methods represented in training, RTs are predicted directly; for methods absent from training, the pretrained backbone is frozen and a new RT head is fine-tuned on a small number ($K$) of measured RTs from the target method (few-shot fine-tuning).
The model is evaluated under three settings with distinct molecule\,$\times$\,method distribution shifts: Molecule-OOD, which evaluates generalization to unseen molecules under chromatographic methods represented in training; Method-OOD, which evaluates generalization to unseen chromatographic methods for molecules largely covered by the training data; and Joint-OOD, which evaluates generalization to both unseen molecules and unseen chromatographic methods.}
\label{fig:concept}
\end{figure}

We show that simulation-derived supervision substantially improves transfer to molecules outside the experimental pretraining domain. Moreover, downstream RT-prediction error decreases systematically as the amount of simulation data increases, exhibiting a significant power-law relationship. This Sim2Real scaling behavior suggests that model transferability can be progressively enhanced by expanding computational data, whose scale is constrained primarily by available computational resources rather than experimental throughput. We further characterize when simulation-derived supervision improves transferability under molecular and chromatographic distribution shifts and benchmark a pre-registered model configuration against from-scratch Graphormer-RT and Uni-RT baselines. Together, these results establish a scalable strategy for complementing finite experimental RT datasets with computational supervision to improve RT prediction beyond the experimental training domain. The overall framework and the three evaluation settings are summarized in Figure~\ref{fig:concept}.

\section{Materials and Methods}

\subsection{Datasets}

Two data sources were used: RepoRT, a public multi-condition HPLC repository~\cite{report2024}, which provides RT-prediction supervision; and PolyOmics, an MD-derived molecular property database generated using the RadonPy pipeline~\cite{polyomics2025,hayashi2022radonpy}, which provides auxiliary supervision.

\subsubsection{RepoRT curation}

Each RepoRT dataset corresponds to a single chromatographic method---defined by a specific column, mobile phase, and gradient program---and is assigned a four-digit identifier, which we use throughout to denote individual methods. Following the reversed-phase selection criteria used in the Graphormer-RT study~\cite{stienstra2025graphormerrt}, we retained methods that (i) used strictly reversed-phase (RP) chromatography, (ii) were operated at a constant flow rate, (iii) used a binary (two-solvent) mobile phase, and (iv) contained at least 50 RT measurements with a valid gradient description. We introduced an additional criterion specific to this study: (v) molecules in a dataset share more than 50\% of molecules in PolyOmics. This criterion ensures sufficient molecular overlap between the experimental RT and auxiliary simulation datasets to provide a meaningful auxiliary supervision signal during training. Importantly, this overlap criterion was applied only to the construction of the training cohort and was deliberately not imposed on the out-of-distribution (OOD) evaluation methods. The Joint-OOD methods described below were instead selected to assess generalization beyond the molecular coverage of the training data.

These filters yielded 179 training methods, with two additional RepoRT methods (0097 and 0238) withheld entirely from training for Method-OOD evaluation. Each method defines a supervised RT-prediction task. Full attrition counts, lists of methods retained at each filtering stage, and the chromatographic column composition of the training cohort are provided in the Supporting Information (Tables~S1 and~S2).

\subsubsection{Simulation auxiliary data}

The auxiliary supervision implements a general strategy for learning highly generalizable molecular representations using molecular simulation data. We used the small-molecule subset of the PolyOmics database~\cite{polyomics2025}. Although PolyOmics and its underlying RadonPy simulation engine~\cite{hayashi2022radonpy} were originally developed for polymer informatics, the present study uses only simulation data for small-molecule compounds.

For each small molecule, a periodic bulk structure consisting of identical molecules was generated and equilibrated using all-atom classical MD simulations. Equilibrium and non-equilibrium MD simulations were then used to estimate a broad range of molecular and bulk properties, including thermal, optical, and mechanical properties, solubility parameters, and structural descriptors. In addition, density-functional theory (DFT) calculations were performed to obtain molecular electronic properties, including the highest occupied molecular orbital (HOMO) and lowest unoccupied molecular orbital (LUMO) energies, dipole moment, and so on. In total, a dataset comprising 39 properties of 21,431 molecules (see Table~S12 for more details) was used as auxiliary supervision. These simulation-derived properties provide diverse, physically grounded supervision for learning molecular representations beyond the chemical space covered by the experimental RT data.

The simulation data were used exclusively as auxiliary molecular-property supervision rather than as direct supervision for chromatographic retention. Their role is to enrich the shared molecular representation with physically grounded information that can enhance transferability across chemical space. Importantly, the chemical-space coverage of this auxiliary supervision can be systematically expanded by generating additional simulation data, independently of the availability of experimental RT measurements. The full simulation protocol is provided in the Supporting Information.

\subsubsection{Evaluation settings}

\begin{figure}[htbp]
\centering
\includegraphics[width=\linewidth]{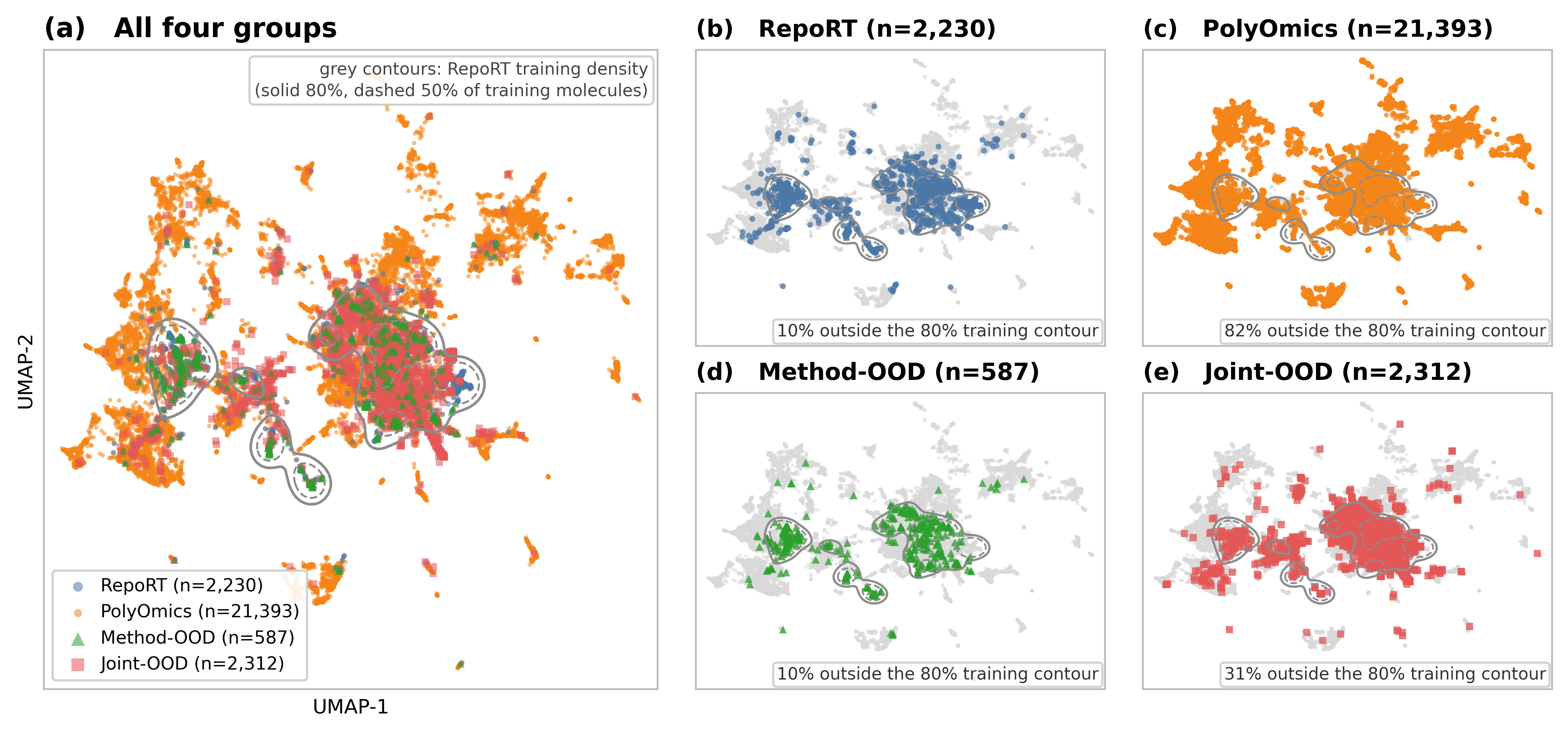}
\caption{\textbf{UMAP visualization of the chemical space of the four molecular groups.}
UMAP was constructed using the ECFP6 fingerprints~\cite{rogers2010ecfp} with the Jaccard metric; the number of molecules in each group is indicated in the panel titles. Individual groups are highlighted against the other groups shown in grey. Grey contours in every panel show the density of the RepoRT training molecules (solid: the contour enclosing 80\% of their density; dashed: 50\%), and each panel states the percentage of its molecules that lie outside the 80\% contour.
\textbf{(a)} Overlay of the four groups shown individually in \textbf{(b--e)}.
\textbf{(b)} RepoRT training molecules.
\textbf{(c)} Molecules in PolyOmics, which span the broadest region of chemical space (82\% outside the 80\% training contour).
\textbf{(d)} Molecules from the Method-OOD datasets, which predominantly occupy well-populated regions similar to those covered by the RepoRT training data (10\% outside the 80\% training contour, the same fraction as the training molecules themselves).
\textbf{(e)} Molecules from the Joint-OOD datasets, many of which are concentrated in and around regions less densely covered by the training data: 31\% lie outside the 80\% training contour, three times the fraction for the training molecules and for Method-OOD, consistent with their low molecular overlap with the training set.}
\label{fig:umap}
\end{figure}

We evaluated model generalization under three complementary settings defined along two independent axes of distribution shift: molecular identity and chromatographic methods (Table~\ref{tab:datasets}). We refer to these settings as \emph{Molecule-OOD}, \emph{Method-OOD}, and \emph{Joint-OOD}. Molecular distribution shift was quantified by the fraction of unique molecules in each evaluation set that appeared in the 179-method training pool, whereas method-level distribution shift was assessed component by component, i.e., by whether the chromatographic column (product and geometry), the eluent system (organic modifier and additives), and the gradient steepness of each evaluation method were represented in the training pool (Table~S2). The chemical spaces of the molecule groups were visualized using UMAP~\cite{mcinnes2018umap} based on ECFP6 fingerprints~\cite{rogers2010ecfp} (Figure~\ref{fig:umap}).

The \emph{Molecule-OOD} setting evaluates generalization to unseen molecules under chromatographic methods represented during training. Molecules within the 179 training methods were partitioned such that no molecular structure appeared in both the training and test splits, resulting in zero molecular overlap by construction while retaining the same chromatographic methods.

The \emph{Method-OOD} setting evaluates generalization to chromatographic methods withheld entirely from training while maintaining high molecular coverage. RepoRT datasets 0097 and 0238 were withheld from training and used exclusively for this evaluation because their molecules are almost entirely represented in the training pool and their chromatographic columns (Waters T3-type C18 columns of commonly used dimensions) are well represented in the training data. Their eluent systems (methanol or acetonitrile with 0.1\% formic acid) and gradient steepness are likewise shared with 64 and 82 training methods, respectively, so that every component of these methods is represented in training and only the specific method---the particular combination of column, eluent, and gradient program measured in a different laboratory---is new (Table~S2). This setting therefore evaluates method-level generalization while minimizing molecular distribution shift.

The \emph{Joint-OOD} setting evaluates simultaneous generalization to both unseen molecules and unseen chromatographic methods. From the latest version of RepoRT datasets, we selected four (0411, 0419, 0420, and 0437) that combine little molecular overlap with the training pool ($\sim$2\%) with chromatographic column models absent from the training data, including some with atypical bore diameters; two of them additionally use eluent additives absent from training (ammonium formate in 0411; heptafluorobutyric acid in 0437) and two use gradient slopes at the extremes of the training range (0420 and 0437; Table~S2). These datasets therefore provide a stringent test of generalization under combined molecular and method distribution shifts.

\subsubsection{Preprocessing}

Registered molecular structures were validated with RDKit~\cite{rdkit} and keyed by InChIKey; invalid records were removed following the dataset guidelines. Retention times are recorded in seconds and used directly, without subtracting the column dead time to convert them to retention factors and without rescaling times across methods. Molecules enter the model as molecular graphs, with no hand-crafted descriptors (for example, logP) appended. FUSE-RT therefore predicts absolute retention time on each method's own time axis, so the mean absolute error (MAE) we report in seconds refers to the raw instrument scale rather than a normalized one. This is possible because the method-specific RT head absorbs each method's dead time and gradient offset, which removes the need for explicit dead-time subtraction or cross-method normalization.

\begin{table}[htbp]
\centering
\small
\caption{\textbf{Three evaluation settings for FUSE-RT.} ``Molecule overlap’’ denotes the fraction of unique molecules in each evaluation set that are also present in the 179-method training pool. For the Molecule-OOD test, the overlap is $0$ by construction, as molecules are partitioned into mutually exclusive training and test sets within the 179 training methods. For the Method-OOD and Joint-OOD settings, molecule overlap is calculated against the pool of training molecules at the exact (full-InChIKey) level. Method identifiers correspond to the four-digit IDs in the RepoRT datasets. Detailed chromatographic column, eluent, and gradient information for each method, together with the number of training methods sharing each component, is provided in Table~S2, and the corresponding chemical-space distributions are shown in Figure~\ref{fig:umap}.}
\label{tab:datasets}
{\small
\begin{tabular}{@{}L{2.0cm}L{3.1cm}L{2.9cm}L{4.1cm}L{2.5cm}@{}}
\toprule
Setting & RepoRT datasets & Molecule overlap with training pool & Method components vs.\ training & Novelty \\
\midrule
Molecule-OOD & 179 (molecule-disjoint test split) & $0\%$ (disjoint split) & same methods & molecules only \\
Method-OOD & 0097, 0238 (2 methods) & $\sim$100\% & column, eluent, and gradient all represented & the specific method (mild) \\
Joint-OOD & 0411, 0419, 0420, 0437 (4) & $\sim$2\% pooled (0.6--2.5\% per method) & columns absent; unseen additives (2 of 4); extreme gradient slopes (2 of 4) & molecules \emph{and} methods \\
\bottomrule
\end{tabular}}
\end{table}

\subsection{Model Architecture}

\begin{figure}[htbp]
\centering
\includegraphics[width=0.95\linewidth]{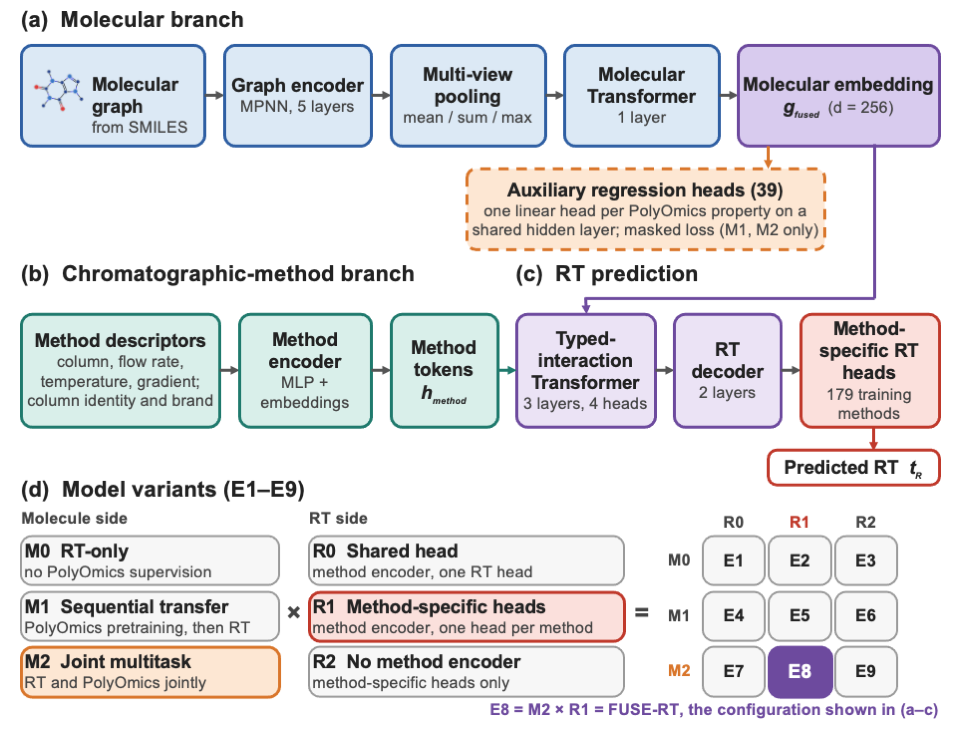}
\caption{\textbf{Architecture of FUSE-RT and the nine model variants.}
\textbf{(a)} Molecular branch: a SMILES-derived molecular graph is encoded by a five-layer message-passing graph encoder, multi-view graph pooling (mean, sum, and max), and a single-layer molecular Transformer, yielding the molecular embedding $g_{\mathrm{fused}}$ ($d_{\text{model}}=256$). When auxiliary supervision is applied (M1 and M2), 39 regression heads (one per PolyOmics property) operating on $g_{\mathrm{fused}}$ predict the PolyOmics properties with a masked regression loss.
\textbf{(b)} Chromatographic-method branch: continuous column and operating descriptors (column length, internal diameter, particle size, flow rate, temperature, and gradient descriptors) together with the categorical column identity and brand are encoded into method tokens $h_{\mathrm{method}}$.
\textbf{(c)} RT prediction: a three-layer typed-interaction Transformer integrates the molecular embedding with the method tokens, a two-layer RT decoder produces the RT representation, and a method-specific head selected according to the chromatographic-method identity outputs the predicted RT $t_R$.
\textbf{(d)} Model variants: three molecular-representation strategies (M0, RT-only; M1, sequential transfer with PolyOmics pretraining followed by RT fine-tuning; M2, joint multitask training on RT and PolyOmics properties) are combined with three RT-prediction architectures (R0, method encoder with a single shared RT head; R1, method encoder with method-specific RT heads; R2, method-specific RT heads without the method encoder), giving the nine variants E1--E9. The architecture shown in \textbf{(a--c)} is E8 (M2\,$\times$\,R1), the default FUSE-RT configuration used throughout.}
\label{fig:architecture}
\end{figure}

The model maps a molecular graph and chromatographic method descriptors to a RT (Figure~\ref{fig:architecture}). A detailed schematic of the same architecture, showing the internal structure of each module, is provided in the Supporting Information (Figure~S3). A five-layer message-passing neural network (MPNN)~\cite{gilmer2017mpnn}, followed by a single-layer molecular Transformer~\cite{vaswani2017attention}, produces a molecular embedding with dimension $d_{\text{model}}=256$. Each chromatographic method is represented by a metadata vector comprising fields indicating continuous column and operating parameters (column length, internal diameter, particle size, flow rate, temperature, and gradient descriptors), together with categorical variables representing column identity and column brand. These features are encoded as
\begin{equation}
\mathbf{h}_{\text{method}}
=
\mathrm{MLP}\!\left(
\left[
\mathbf{x}^{\text{cont}}_{\text{col}}
\;\|\;
\mathrm{Emb}(\mathbf{x}^{\text{cat}}_{\text{col}})
\;\|\;
\mathrm{Emb}(\mathbf{x}^{\text{cat}}_{\text{brand}})
\right]
\right),
\end{equation}
where the brand vocabulary is constructed using the training methods only. A three-layer typed-interaction Transformer with four attention heads integrates the molecular embedding with the chromatographic-method representation, and a two-layer decoder maps the resulting representation to the predicted RT. When auxiliary supervision is applied, a set of 39 regression heads (one per PolyOmics property; Table~S12) operating on the molecular embedding predicts the PolyOmics properties, with missing target values handled using a masked regression loss.

\subsection{Model Variants and Training}

We evaluated nine model variants by combining three molecular-representation learning strategies (M0--M2) with three RT-prediction architectures (R0--R2) (Table~\ref{tab:matrix} and Figure~\ref{fig:architecture}). The molecular-representation strategies were defined as follows: \textbf{M0 (RT-only)}, trained without PolyOmics auxiliary supervision; \textbf{M1 (sequential transfer)}, pretrained on PolyOmics-property prediction and subsequently fine-tuned on RT prediction~\cite{hu2020pretraingnn}; and \textbf{M2 (joint multitask)}, jointly trained on RT and PolyOmics-property prediction tasks. The RT-prediction architectures were defined as follows: \textbf{R0 (shared head)}, using a chromatographic-method encoder and a single RT-prediction head shared across all methods; \textbf{R1 (method-specific heads)}, using a chromatographic-method encoder together with method-specific RT-prediction heads; and \textbf{R2 (method-specific heads without method encoding)}, using method-specific RT-prediction heads without explicit chromatographic-method input. Combining M0--M2 with R0--R2 yielded nine model variants, labeled \textbf{E1--E9}.

Each of the 179 training methods was split at the molecule level. All records sharing the same full InChIKey were assigned exclusively to the training, validation, or test set, thereby preventing exact molecular structures from appearing across different datasets. The absence of exact-structure leakage was verified for all ten random seeds. The same ten molecule-level splits were used for all nine model variants, enabling paired comparisons across models. Models were trained using AdamW~\cite{loshchilov2019adamw} with a learning rate of $2\times10^{-4}$, weight decay of $10^{-4}$, gradient clipping at 5.0, and dropout of 0.18. ReduceLROnPlateau scheduling and early stopping based on the validation score were applied. Full hyperparameter settings are provided in Table~S14.

For adaptation to an unseen chromatographic method, the pretrained backbone was frozen and only a newly initialized two-layer RT prediction head was trained using $K$ additional target data, which were kept disjoint from the query set~\cite{kumar2022finetuning}. At $K=0$ (zero-shot prediction without transfer), R0 directly uses its shared RT head. Because R1 and R2 contain only method-specific heads and therefore have no prediction head corresponding to an unseen method, their zero-shot predictions were obtained by averaging predictions from the heads of the training methods. Thus, zero-shot comparisons among R0--R2 reflect their respective prediction protocols rather than a direct ablation of the RT-head architecture. For Joint-OOD evaluation, we used $K\in\{5,20,100,1000\}$, whereas for Method-OOD evaluation we used $K\in\{5,20,100,200\}$; method 0097 permitted evaluation only up to $K=20$.

Predictive performance was evaluated using the mean absolute percentage error (MAPE), coefficient of determination ($R^2$), MAE, root mean squared error (RMSE), and Spearman's rank correlation coefficient. MAE and RMSE are reported in seconds (s). Because MAPE can be disproportionately affected by analytes with short RTs~\cite{hyndman2006accuracy}, MAE and $R^2$ are reported alongside it to provide complementary measures of predictive performance. Statistical comparisons between models were performed using paired, two-sided $t$-tests on the per-seed differences in each metric across the ten shared molecule-level splits. For OOD evaluations involving multiple chromatographic methods, metrics were macro-averaged across methods to prevent methods with larger sample sizes from disproportionately influencing the overall performance.

\begin{table}[htbp]
\centering
\small
\caption{\textbf{The nine model variants defined by combinations of molecular-representation strategies and RT-prediction architectures.} Rows represent the molecular-representation strategies (M0--M2), and columns represent the RT-prediction architectures (R0--R2); each cell denotes one model variant (E1--E9). R0 and R1 incorporate chromatographic metadata through a chromatographic-method encoder, whereas R2 does not use explicit chromatographic-method input. For transfer learning to an unseen chromatographic method, R0 enables zero-shot prediction through its shared RT head, whereas R1 and R2, which use method-specific RT heads, are adapted by fitting a newly initialized head using $K$ target data.}
\label{tab:matrix}
\begin{tabular}{@{}L{4.4cm}C{3.5cm}C{3.5cm}C{3.5cm}@{}}
\toprule
\textbf{Molecular representation}
& \textbf{R0}: Method encoder
& \textbf{R1}: Method encoder
& \textbf{R2}: No method encoder \\
& \textbf{Shared RT head}
& \textbf{Method-specific RT heads}
& \textbf{Method-specific RT heads} \\
\midrule
\textbf{M0}: No auxiliary supervision   & E1 & E2 & E3 \\
\textbf{M1}: PolyOmics pretraining      & E4 & E5 & E6 \\
\textbf{M2}: PolyOmics multitask learning & E7 & E8 & E9 \\
\bottomrule
\end{tabular}
\end{table}

\section{Results and Discussion}

\subsection{Simulation Supervision in the Molecule-OOD Setting}

We evaluate generalization in the Molecule-OOD setting, where chromatographic methods are represented during training but the test molecules are unseen. Among the nine model variants, E8 (M2: joint RT $+$ PolyOmics multitask learning; R1: chromatographic-method encoder with method-specific RT heads) achieves the lowest MAPE of $19.22\pm2.64\%$ and an $R^2$ of $0.929$ across the ten molecule-level splits. E5 (M1: PolyOmics pretraining followed by RT training; R1) also achieves statistically comparable performance. Auxiliary simulation supervision provides a modest but consistent improvement in this setting. Because all models were evaluated on the same ten molecule-level splits, statistical comparisons were performed using paired tests. E8 improves over its simulation-free counterpart E2 by $1.15$ percentage points in MAPE (paired two-sided $t$-test, $p=0.044$), while E1, which uses neither simulation supervision nor method-specific heads, has a MAPE $2.81$ percentage points higher than E8 ($p<0.001$). Complete results for all model variants and paired statistical comparisons are provided in the Supporting Information (Tables~S3 and~S4).

\begin{figure}[htbp]
\centering
\includegraphics[width=\linewidth]{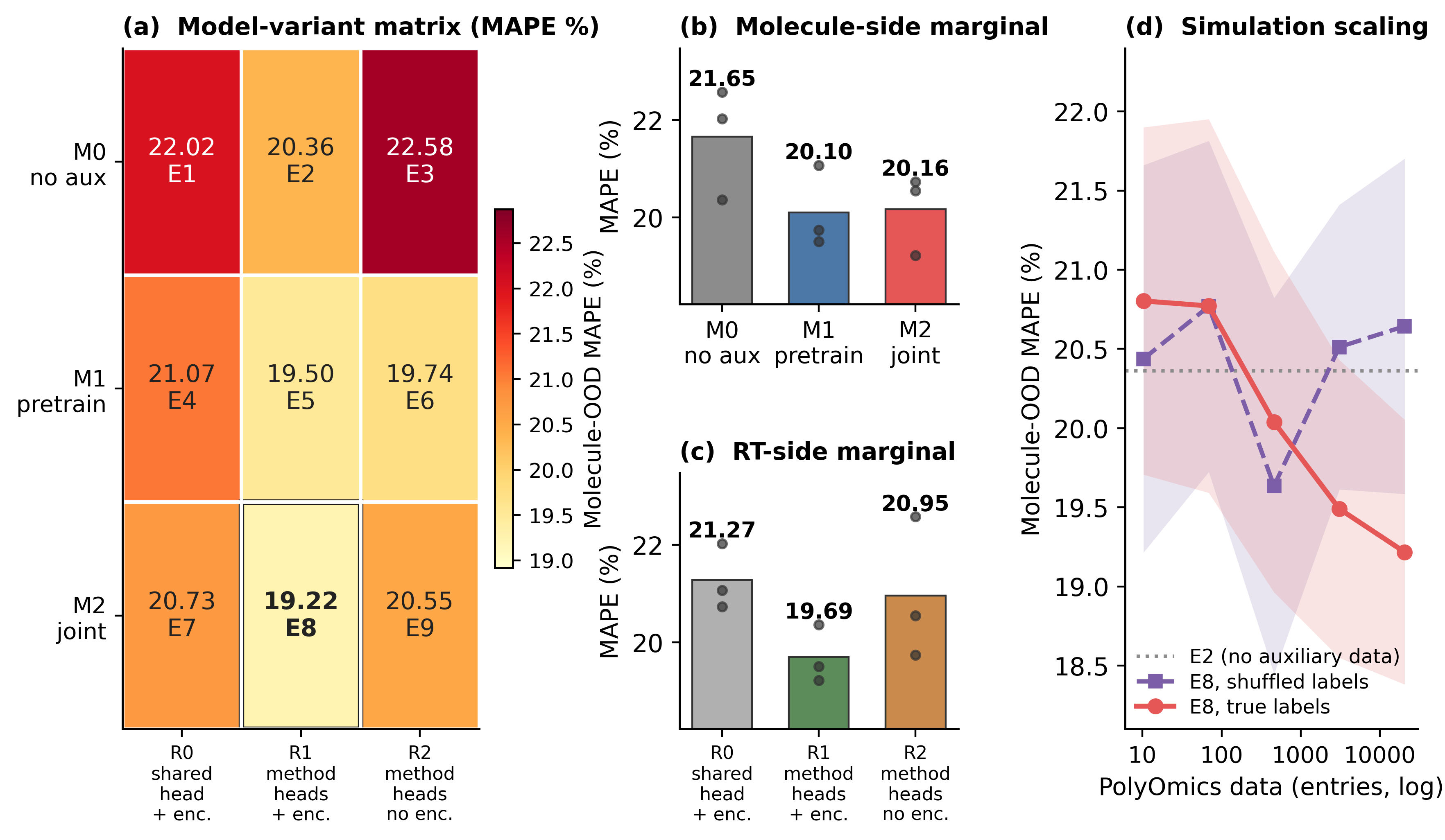}
\caption{\textbf{Performance and simulation-data scaling in the Molecule-OOD setting.} Models are evaluated on the Molecule-OOD test sets comprising unseen molecules under chromatographic methods represented during training. \textbf{(a)} MAPEs (\%) for the nine model variants defined by combinations of the molecular-representation strategies (M0--M2) and RT-prediction architectures (R0--R2); E8 was the best-performing variant. \textbf{(b)} Marginal predictive performance of the three molecular-representation strategies and \textbf{(c)} the three RT-prediction architectures. Bars indicate the mean MAPEs, and dots represent the three constituent model variants. PolyOmics auxiliary supervision (M1 and M2) and the chromatographic-method encoder with method-specific RT heads (R1) reduce prediction error. \textbf{(d)} Scaling of MAPEs with the amount of PolyOmics data (number of simulation entries) used for auxiliary supervision in the joint model E8. The $x$-axis is logarithmically scaled over approximately three orders of magnitude, and shaded bands indicate $\pm1$ SEM across ten seeds; the simulation-free model E2 is shown as a dotted reference. MAPE decreases systematically as the simulation dataset increases, whereas random permutation of the property values across molecules eliminates this scaling behavior and yields performance comparable to the simulation-free reference. This control indicates that the improvement arises from the molecular information encoded in the simulated properties rather than merely from introducing additional prediction tasks. All comparisons use the same ten molecule-level splits; paired statistical analyses are provided in the Supporting Information.}
\label{fig:ablation}
\end{figure}

The systematic comparison of the nine model variants further allows the effects of the molecular-representation strategy and RT-prediction architecture to be examined separately (Figure~\ref{fig:ablation}). Averaged across the three RT-prediction architectures, PolyOmics supervision reduces the mean MAPE from $21.65\%$ for M0 (RT-only) to $20.10\%$ for M1 (sequential transfer) and $20.16\%$ for M2 (joint multitask learning). No significant difference is observed between M1 and M2, indicating that sequential pretraining provides comparable performance to joint multitask learning in the Molecule-OOD setting. Averaged across the three molecular-representation strategies, R1, which combines the chromatographic-method encoder with method-specific RT heads, achieves the lowest mean MAPE ($19.69\%$), compared with R0 ($21.27\%$), which uses a shared RT head, and R2 ($20.96\%$), which omits explicit chromatographic-method encoding. These results indicate that, for chromatographic methods represented during training, combining explicit method information with method-specific RT heads provides the most accurate predictions.

\subsection{Scaling with Simulation Data}

Simulation supervision is intended to extend molecular representations beyond the coverage of experimental RT data. We examined whether increasing the amount of simulation data systematically improves RT prediction. Figure~\ref{fig:ablation}(d) shows the Molecule-OOD test MAPE as a function of the amount of PolyOmics data (the number of simulation entries) used for joint training, spanning approximately three orders of magnitude. Under the same training protocol used for the E1--E9 model variants, prediction error decreases systematically as the simulation dataset increases. The observed trend is consistent with Sim2Real scaling previously reported for Sim2Real transfer in materials and chemical modeling~\cite{mikami2023syn2real,minami2025sim2real,aoki2023miscibility,yamada2019shotgun}, as well as with broader neural scaling laws for pretraining and transfer~\cite{kaplan2020scaling,hernandez2021scalingtransfer}. Unlike experimental RT measurements, simulation data can be systematically expanded using computational resources, both in the number and diversity of molecules and in the range of calculated properties.

To determine whether this scaling arises from the information encoded in the simulated properties rather than from a generic regularization effect of auxiliary learning, we conducted the scaling analysis by randomly permuting each PolyOmics property across molecules. This procedure preserves the marginal distributions of the simulated properties while disrupting their associations with molecular structures. All other factors, including the molecules, auxiliary targets, loss function, model architecture, and training protocol, were kept unchanged. In contrast to the original model, the shuffled-property control exhibits no systematic improvement or scaling behavior (Figure~\ref{fig:ablation}d). Across the ten shared molecule-level splits, the original model shows an improvement of $0.54\pm0.14$ MAPE percentage points per decade of PolyOmics data (one-sample $t$-test against zero, $p=0.004$), whereas the shuffled-property control shows no significant scaling ($+0.02\pm0.16$ percentage points per decade, $p=0.91$); the difference between the slopes is significant (paired $t$-test, $p=0.025$). At the full simulation-data scale, the original model achieves a MAPE of $19.22\%$, compared with $20.64\%$ for the shuffled-property control (paired difference, $1.43$ percentage points; $p=0.025$). Moreover, the original model significantly outperforms the no-auxiliary E2 reference ($20.36\%$ MAPE; paired $p=0.044$), whereas the shuffled-property control does not ($p=0.37$). Together, these results indicate that the observed scaling arises from molecular information encoded in the simulated property datasets rather than from the mere addition of auxiliary prediction tasks.

\subsection{Generalization to Out-of-Distribution Chromatographic Methods}
\label{sec:ood}

\begin{figure}[htbp]
\centering
\includegraphics[width=\linewidth]{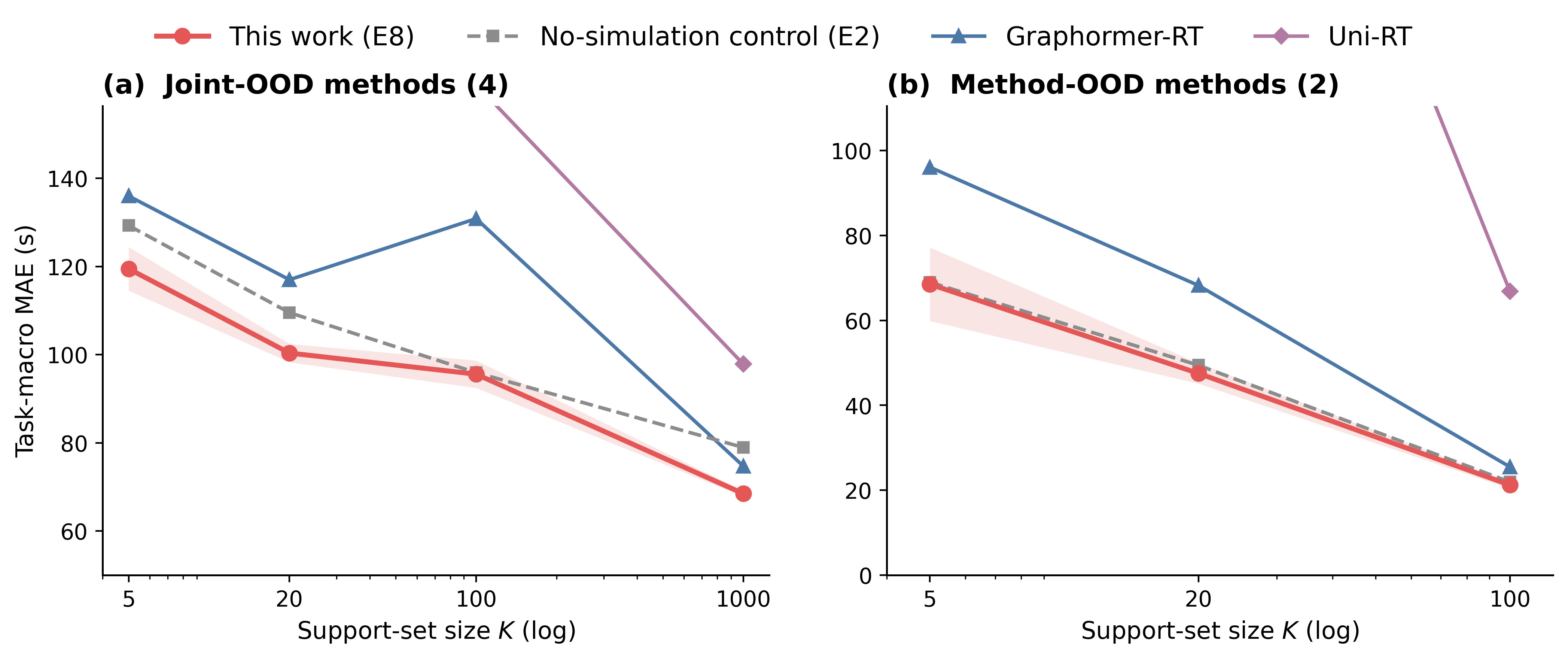}
\caption{\textbf{$K$-shot adaptation in the out-of-distribution settings.}
Task-macro-averaged MAE (s) is shown as a function of training data size $K$ for four pre-registered model configurations: E8 (this work), its simulation-free control E2, Graphormer-RT, and Uni-RT. The $x$-axis shows $K$ on a logarithmic scale, and the $y$-axis shows MAE. \textbf{(a)} Joint-OOD evaluation on four methods with low molecular overlap and chromatographic methods outside the training distribution. \textbf{(b)} Method-OOD evaluation on two withheld methods close to the training distribution. E8 and E2 use method-specific RT heads and therefore require additional training data, with evaluation starting at $K=5$. For readability, the $y$-axis range is restricted to the range of E8, E2, and Graphormer-RT; consequently, Uni-RT falls outside the displayed range at small $K$ and re-enters the plotted range as $K$ increases. The number of methods contributing to the task-macro average decreases at larger $K$ because some datasets are too small to provide the required training datasets: for Joint-OOD, $n_t=4,4,3,1$ at $K=5,20,100,1000$, respectively; for Method-OOD, $n_t=2,2,1$ at $K=5,20,100$, respectively. E8 achieves lower or comparable errors relative to E2 and the baseline models across the Joint-OOD setting, whereas E8 and E2 show similar performance in the Method-OOD setting. Shaded bands around the E8 curves denote $\pm1$ SEM. Individual method-level curves are provided in Figure~S2, and detailed results for each model configuration and baseline are reported in Tables~S5, S6, and~S10.}
\label{fig:kshot}
\end{figure}

The central question is how well the model generalizes to chromatographic methods absent from training. To this end, we examine two contrasting OOD settings (Table~\ref{tab:datasets}; Figure~\ref{fig:umap}). The two Method-OOD evaluation datasets (0097 and 0238) provide a near-in-distribution reference: their molecules ($\sim$100\% overlap) and chromatographic columns (Waters T3-type C18 columns with commonly used dimensions) are both well represented in the training data, as are their eluent systems and gradient slopes (Table~S2), such that the unseen chromatographic method is the primary source of domain shift. In contrast, the four Joint-OOD datasets (0411, 0419, 0420, and 0437) contain structurally novel molecules (only $\sim$2\% overlap) and use chromatographic columns absent from the training data, two of them also using eluent additives absent from training and two using gradient slopes at the extremes of the training range, thereby requiring simultaneous generalization across molecular and chromatographic-method domains. Comparing these two settings allows us to examine how transfer performance changes as the coverage gap extends from chromatographic methods alone to both molecules and methods. In both settings, adaptation freezes the entire backbone and fits a newly initialized two-layer RT head using $K$ support molecules and their RT measurements. Figure~\ref{fig:kshot} reports the task-macro-averaged MAE across the methods in each setting as a function of the number of support molecules; individual method-level curves are provided in the Supporting Information (Figure~S2).

For the Method-OOD tests (Figure~\ref{fig:kshot}b), prediction errors are relatively low and adaptation is effective with limited target data. Because the molecular space is already well covered by the training data, the shared RT head enables prediction without target-method labels; for example, task 0238 achieves a zero-shot MAPE of $9.4\%$ ($R^2=0.95$). The E8 model and its simulation-free counterpart E2 show similar performance, indicating that simulation supervision provides little additional benefit when the relevant molecular space is already well represented. This setting also highlights an architectural distinction that becomes important for unseen chromatographic methods: the shared-head architecture (R0) can directly generate zero-shot predictions for a new method, whereas the method-specific-head architectures (R1 and R2) have no corresponding head for an unseen method and therefore require target data to additionally train a new head. R0 is thus better suited to zero-shot transfer, whereas R1 achieves the highest accuracy for methods represented during training.

The Joint-OOD tests exhibit a markedly different behavior (Figure~\ref{fig:kshot}a). Zero-shot prediction performs poorly across all models, with MAEs of $260$--$325$\,s and strongly negative $R^2$ values (Table~\ref{tab:baseline}), indicating that the RT of an unseen chromatographic method is difficult to infer without fine-tuning. However, performance improves rapidly with few-shot adaptation, and the benefit of simulation supervision becomes more pronounced in this setting. E8 performs at or below the error of its simulation-free counterpart E2 across nearly all support sizes and methods, with the largest improvements observed in the small-target-data regime. For example, at $K=100$, the task-macro-averaged MAPE is approximately $19$--$20\%$ for the PolyOmics-supervised model, compared with approximately $25\%$ for its simulation-free counterpart.

Importantly, the $>50\%$ PolyOmics-overlap criterion in molecules was applied only when constructing the 179-method training cohort and was deliberately not imposed on the Joint-OOD methods. Consequently, molecules in the Joint-OOD tests have only $\sim$6\% overlap with the PolyOmics simulation dataset. The observed improvement therefore cannot be explained simply by direct simulation coverage of the target molecules, but instead supports the transfer of molecular representations learned from simulation supervision to previously uncovered chemical space. Because the backbone is frozen throughout the adaptation, these gains reflect the quality of the transferable representation rather than task-specific learning. Individual method-level curves are provided in Figure~S2, and detailed results for each method and model configuration are reported in Tables~S5 and~S6.

\subsection{Benchmarking against Existing Models}

We benchmark our models against two recent RT-prediction models, Graphormer-RT~\cite{stienstra2025graphormerrt,ying2021graphormer} and Uni-RT~\cite{xiong2026unirt}. We focus the comparison primarily on the Joint-OOD setting rather than on the 179 training methods because the two baselines were developed using different training datasets and protocols. Comparison under a common task involving adaptation to unseen chromatographic methods and molecules therefore provides a more relevant assessment of the generalization capability central to this study. Each baseline was evaluated in a manner consistent with its original design. Uni-RT was trained on a separately curated subset of RepoRT using its own preprocessing criteria and separate models for reversed-phase and HILIC modes. Nevertheless, its reversed-phase training set is similar to ours in terms of unique-molecule coverage: the two datasets overlap substantially and contain comparable numbers of unique molecules (Supporting Information, Table~S16). Thus, differences in performance between Uni-RT and our model are unlikely to be attributable solely to differences in the molecular space represented during training, but may also reflect differences in data curation, training strategy, and model architecture. Because retraining Uni-RT on our 179-method dataset would deviate from its original per-mode training design, we directly evaluate the released pretrained model. Graphormer-RT, in contrast, was originally evaluated using a random split that allows the same molecules to appear in both the training and test sets, making direct assessment of molecular generalization under that protocol inappropriate for the present comparison. We therefore retrained Graphormer-RT from scratch on our 179 methods using the same molecule-disjoint splits as those used for our model and the released code.

Under the common frozen-backbone $K$-shot adaptation protocol (Table~\ref{tab:baseline}; Figure~\ref{fig:kshot}), E8 achieves the lowest MAE and highest $R^2$ at every target data size evaluated in the Joint-OOD setting. At $K=20$, E8 achieves an MAE of $100.4$\,s ($R^2=0.27$), compared with $117.0$\,s ($R^2=0.12$) for Graphormer-RT and $212.9$\,s (with a strongly negative $R^2$) for Uni-RT, although the performance gap narrows as the amount of target data increases. In the Molecule-OOD setting, E8 also achieves an MAE of $36.0$\,s ($R^2=0.93$), compared with $64.9$\,s ($R^2=0.80$) for the retrained Graphormer-RT, and yields a lower MAE across all ten molecule-level splits.

\begin{table}[htbp]
\centering
\small
\caption{\textbf{Baseline comparison in the Joint-OOD setting.}
Results are reported for the four Joint-OOD tests as task-macro averages across the $n_t$ methods for which a training dataset of size $K$ is available. Performance is evaluated using MAE (s) and $R^2$. All models are evaluated as single, pre-registered configurations using the same frozen-backbone $K$-shot adaptation protocol; our model corresponds to E8. A zero-shot ($K=0$) reference is also included. Because E8 uses method-specific RT heads and therefore has no prediction head for an unseen method, the shared-head configuration E7 (M2$\times$R0) is used for our zero-shot result ($\dagger$). The number of contributing methods ($n_t$) decreases at larger $K$ because some datasets are too small to provide the required training dataset; at $K=1000$, only method 0437 contributes. Bold values indicate the best-performing model at each $K\geq5$.}
\label{tab:baseline}
\fittable{%
\begin{tabular}{@{}cccccccc@{}}
\toprule
& & \multicolumn{3}{c}{MAE (s)} & \multicolumn{3}{c}{$R^2$} \\
\cmidrule(lr){3-5}\cmidrule(lr){6-8}
$K$ & $n_t$ & This work (E8) & Graphormer-RT & Uni-RT & This work (E8) & Graphormer-RT & Uni-RT \\
\midrule
$0^{\dagger}$ & 4 & $298.3$ & $321.4$ & $261.3$ & $-4.26$ & $-5.26$ & $-1.72$ \\
5    & 4 & $\mathbf{119.5}$ & $136.0$ & $413.8$ & $\mathbf{-0.028}$ & $-0.256$ & $-33.52$ \\
20   & 4 & $\mathbf{100.4}$ & $117.0$ & $212.9$ & $\mathbf{0.270}$  & $0.120$  & $-4.529$ \\
100  & 3 & $\mathbf{95.6}$  & $130.9$ & $161.3$ & $\mathbf{0.463}$  & $0.221$  & $-0.414$ \\
1000 & 1 & $\mathbf{68.5}$  & $74.7$  & $97.9$  & $\mathbf{0.811}$  & $0.784$  & $0.634$  \\
\bottomrule
\end{tabular}}
\end{table}

\subsection{Remarks}

Three caveats should be considered when interpreting these results. First, molecular novelty in the Molecule-OOD test is enforced at the level of exact molecular structure, defined by the full InChIKey including stereochemistry and salt form. Thus, no test molecule is identical to a training molecule, although approximately $20\%$ share the same 2D molecular skeleton with a training molecule, for example as a stereoisomer or salt form. The Molecule-OOD test therefore excludes exact structural overlap but not close structural analogues; a skeleton-disjoint split would provide a more stringent assessment of molecular novelty. By contrast, molecules in the Joint-OOD setting share almost no exact structures or 2D molecular skeletons with the training molecules (at most $\sim$4\% skeleton overlap; Table~S15), providing a stricter test of transfer to previously unseen chemical structures.

Second, the Method-OOD evaluation contains only two methods that are relatively close to the training distribution. Consequently, the present assessment of generalization relies primarily on the Molecule-OOD setting for molecular novelty and the Joint-OOD setting for simultaneous molecular and chromatographic-method novelty. Evaluation on a larger and more diverse collection of Method-OOD methods will be needed to characterize method-level generalization more comprehensively.

Third, the baseline models are not fully matched in terms of their original training data and procedures. Graphormer-RT was retrained on our 179-method dataset and therefore uses fewer training records than in its original implementation, whereas Uni-RT was evaluated using its released pretrained model. Accordingly, these comparisons should be interpreted as benchmarks under the present generalization and adaptation settings rather than as definitive comparisons of the maximum achievable performance of the respective architectures.

\section{Conclusions}

We framed the generalization of machine-learned HPLC retention-time prediction as a problem of limited molecular and chromatographic-method coverage in experimental training data and investigated physics-based simulation data---which can be scaled without requiring additional experiments---as a means of extending this coverage. A multitask foundation model jointly supervised by experimental RT data and molecular properties derived from molecular simulations allowed us to examine this hypothesis through a systematic comparison of nine model variants combining three molecular-representation strategies with three RT-prediction architectures. Although PolyOmics was used as the simulation source in this study, the broader premise is that molecular simulation can provide scalable auxiliary supervision beyond the chemical space covered by experimental retention data.

The benefit of auxiliary simulation supervision depends strongly on the coverage gap between the training and target domains. Its contribution is limited in the Molecule-OOD setting and in the two Method-OOD methods, where the relevant molecular space is already well represented in the training data, but becomes more pronounced in the Joint-OOD setting, where both molecules and chromatographic methods differ substantially from those encountered during training. Under few-shot adaptation with a frozen backbone, the simulation-supervised representation generally provides greater benefit in this low-coverage regime, particularly when target data are scarce. The Method-OOD results further show that, when molecular coverage is high and only the chromatographic method is unseen, a shared RT head can enable zero-shot prediction without target-method RT labels. A property-shuffling control provides additional evidence that the observed simulation-data scaling arises from molecular information encoded in the simulated properties: randomly permuting each simulated property across molecules while otherwise preserving the auxiliary-learning setting eliminates the systematic improvement with increasing simulation-data size. The architectural comparison also reveals a trade-off between specialization and transferability. The chromatographic-method encoder with method-specific RT heads provides the highest accuracy for methods represented during training, whereas the shared-head architecture enables direct zero-shot prediction for unseen methods. Sequential simulation pretraining and joint multitask learning provide comparable performance in the Molecule-OOD setting, suggesting that sequential pretraining may offer a simpler alternative when joint training is computationally demanding.

Several directions remain for further investigation. The Method-OOD evaluation should be extended beyond the two relatively near-distribution methods considered here to include a larger and more diverse range of chromatographic conditions, while the Joint-OOD evaluation should likewise be expanded to better characterize simultaneous molecular and method-level distribution shifts. A matched-protocol comparison in which Uni-RT is retrained on the same training data would also enable a more controlled architectural benchmark. More generally, adaptation strategies beyond fitting a newly initialized RT head, such as residual or affine calibration of the retention scale, may improve data efficiency and reduce the risk of performance degradation when the target domain is already well covered by the pretrained representation. The broader implication of this study is that physics-based molecular simulation provides a scalable and underexplored source of auxiliary supervision for retention modeling. Systematically expanding such supervision in chemical diversity and in the range of computed molecular properties, rather than merely increasing data volume, may provide a practical route toward extending the generalization capability of HPLC foundation models into regions of molecular and chromatographic-method space that remain sparsely covered by experimental retention data.

\section{Author Contributions}

All authors contributed to and have given approval for the final version of the manuscript. Wu, S., conceptualization, methodology, software, validation, formal analysis, writing---original draft, writing---review \& editing; Han, Y., software, validation, formal analysis, investigation, data curation; Hayashi, Y., resources, data curation, software, writing---review \& editing; Yoshida, R., conceptualization, methodology, supervision, project administration, funding acquisition, writing---review \& editing; Mito, Y., Watabe, Y., Takaya, H., Kubo, T., investigation, resources, writing---review \& editing.

\section{Conflicts of Interest}

The authors declare no conflicts of interest.

\begin{acknowledgement}

This research was primarily supported by the Japan Science and Technology Agency (JST) (Grant Number JPMJCR2332) and the Ministry of Education, Culture, Sports, Science and Technology (MEXT) through the ``Program for Promoting Researches on the Supercomputer Fugaku'' (JPMXP1020200314) and ``Next-Generation Computational Science Global Reach and Innovation Program'' (JPMXP1920260105). Computational resources were provided by Fugaku at the RIKEN Center for Computational Science, Kobe, Japan (hp230190, hp240216, hp250235, and hp260407).

\end{acknowledgement}

\section{Data and Code Availability}

The experimental RT measurements were obtained from the public RepoRT repository.\cite{report2024} The PolyOmics simulation dataset used for auxiliary supervision is available at \url{https://huggingface.co/datasets/yhayashi1986/PolyOmics}. The source code for \mbox{FUSE-RT}, the trained model weights, and the scripts required to reproduce the results reported here are available at \url{https://github.com/jueminghaha/FUSE-RT}.

\begin{suppinfo}
RepoRT curation attrition (Table~S1); column composition of the training and Method-OOD/Joint-OOD methods (Table~S2); full Molecule-OOD results and paired comparisons for all nine models (Tables~S3, S4); Method-OOD per-method results (Table~S5, Figure~S1) and full per-model matrices (Tables~S7--S9); Joint-OOD per-configuration results and per-method adaptation curves (Table~S6, Figure~S2); baseline comparison details (Table~S10); simulation-data scaling (Table~S11); the small-molecule simulation protocol; model, feature, and training details, including auxiliary targets, chromatographic-condition fields, and hyperparameters (Tables~S12--S14); the molecule-disjointness and overlap audit (Table~S15); and the Uni-RT reversed-phase training-set overlap (Table~S16).
\end{suppinfo}

\bibliography{acs_hplc_fm}

\end{document}